\documentclass[%
 preprint,
 amsmath,amssymb,
 aps,
prb,
showkeys]{revtex4-1}

\usepackage{graphicx}
\usepackage{dcolumn}
\usepackage{bm}
\usepackage{lineno}
\usepackage{natbib}

\begin{document}


\title{Risk evaluation and behaviour: defining appropriate frames of reference}
\author{Jared M. Field}
\email{jared.field@maths.ox.ac.uk}
\affiliation{%
Wolfson Centre for Mathematical Biology, Mathematical Institute, University of Oxford, Oxford OX2 6GG,
United Kingdom}%
\altaffiliation[Also at ]{Mathematical Ecology Research Group, Department of Zoology, University of Oxford, Oxford, UK}


\author{Michael B. Bonsall}
\affiliation{Mathematical Ecology Research Group, Department of Zoology, University of Oxford, Oxford, UK}


\date{\today}

\begin{abstract}

Many biological, psychological and economic experiments have been designed where an organism or individual must choose between two options that have the same expected reward but differ in the variance of reward received. In this way, designed empirical approaches have been developed for evaluating risk preferences. Here, however, we show that if the experimental subject is inferring the reward distribution (to optimize some process), they will never agree in finite time that the expected rewards are equal. In turn, we argue that this makes discussions of risk preferences, and indeed the motivations of behaviour, not so simple or straightforward to interpret.  We use this particular experiment to highlight the serious need to consider the frame of reference of the experimental subject in studies of behaviour. 


\end{abstract}

\maketitle
\section*{Introduction}
While in physics it is standard, in biology we do not often think about frames of reference. Suppose, for example, an experiment is set-up such that an organism (or person) must choose between two options with the same expected reward but different variance in reward.  Further, suppose that the organism consistently chooses the safe bet (lower variance).  In this case, we may be tempted to label that organism as risk-averse. This makes sense from our point of view or frame of reference. However, are we sure that the organism agrees on the experimental set-up? Perhaps, from the frame of reference of the organism, the expected rewards are not even perceived to be the same.  If this is the case, we ought to be careful with the conclusions we draw about why the organism is behaving a certain way. Though this particular experiment was picked to make a point, hundreds of variations on it have been carried out across several disciplines (see reviews  \cite{kacelnik1996risky,shafir2000risk, weber2004predicting} and references therein).

Most theoretical attempts to understand behaviour start by assuming some sort of quantity to be optimized \cite{parker1990optimality}.  In finance, classically it is return on investments under a certain risk constraint \cite{markowitz1952portfolio}. In economics, it is  (broadly defined) utility \cite{morgenstern1953theory}. In biology,  utility is usually replaced with reproductive values and assumptions of rationality with natural selection \cite{mcnamara1980application}.  However, energy budgets and threshold reserves have also proved fruitful in understanding risky behaviour\cite{stephens1981logic}. Here,  however, we divorce ourselves of any such quantities.  Instead, we shift attention to the information available in order to make decisions. This way the focus is put on potential differences in beliefs between experimenters and those on whom they are experimenting. We are interested not in predicting traits or behaviours  in experiments but in the conclusions we can and cannot draw from them.

In this paper, as is standard in statistical decision theory, we assume that the organism or person being studied is sampling their environment and updating their beliefs \cite{mcnamara2006bayes,pike2016general,field2018ignorance,Sutton20118}.  In particular, in the experiment described above (and below, in more detail),  we suppose that for each of the options the organism is inferring the probability of receiving a reward at any instance. Our analyses show that, given infinite time, the organism can indeed infer correctly this probability. In other words, the organism will agree with the experimenter on the experimental set-up (which is to say, agree that the expected rewards of each option are equal). However, we then go on to show that in finite time such agreement will never be reached. This is clearly important as all experiments, by nature, have a fixed end-point. In light of this, we point out that it is not so simple to infer risk preferences, as is often done,  from these types experiments.  The problem, as we point out, arises from the experimenter and experimental subject having different frames of reference. 

The rest of the paper is organized as follows: in the next section we go into more detail on the broad class of experiments we are using to demonstrate our point. Following this we introduce the inference problem and prove convergence to the true probability of receiving a given reward given infinite time. Next, noting that real experiments are finite, we show that such a convergence does not occur in this case. Further, we show that the ramification of this is that the experimenter and experimental subject do not agree in finite time on the parameters of the experiments. Otherwise put, for any finite number of trials the organism (or person) being experimented on will not believe that the expected rewards of each option are equal. Finally, we summarise our findings and consider their broader implications.

\section*{Experimental problem}
To motivate our problem, we consider an experimental design where an organism is presented with two choices. One choice (called an arm) leads always (\emph{i.e.}with probability $\hat{p}_1=1$) to a fixed reward of $c$. The other choice (the other arm) leads with probability $\hat p_2$ to a reward of $a$ and $(1-\hat{p}_2)$ to a reward of $b$. The experiment is designed such that the expected reward is the same on both arms. This leads to, 

\begin{equation}
c= \hat{p}_2a+(1-\hat{p}_2)b.
\end{equation}
In this way, from the point of view of the experimenter, if only expected values are used then the organism should be indifferent to each arm. In light of this, in order to make decisions the organism ought to consider the variances in rewards. For an excellent (and extensive) review of these types of experiments see \cite{kacelnik1996risky}. However, as we show below, from the point of view of the organism, the expected rewards are not always equal. 

\section*{Inference problem}
Suppose an organism is attempting to infer the distribution of two possible pay-offs on an experimental arm of the type described above. In particular, by sampling the past pay-offs they are attempting to infer, in a Bayesian manner, the probability $p$ of receiving $g$ and the probability $1-p$ of receiving $h$.  In order to do this, both a prior distribution and likelihood function will be needed to form a posterior distribution for $p$. To this end, suppose $s$ successes and $f$ failures (with regards to receiving $g$) are observed. In this case,  the likelihood that $p=x$ will be given by 
\begin{equation}
Pr(s,f|p=x) = \binom{s+f}{s}x^s(1-x)^f .\label{1.1}
\end{equation}
To encode any previous knowledge we choose the conjugate prior of the binomial distribution, the beta distribution, which is given by:
\begin{equation}
Pr(p=x) = \frac{x^{\alpha -1}(1-x)^{\beta -1}}{B(\alpha, \beta)}, \label{1.2}
\end{equation}
where $B(\alpha, \beta)$ is the beta function with $\beta>0$ and $\alpha>0$. The benefit of using this prior is two-fold. First, being the conjugate prior of \eqref{1.1}, it makes possible the calculation of a closed-form posterior distribution. Second, the hyper parameters $\alpha$ and $\beta$ allow for the encoding of an incredibly wide range of prior beliefs (for example $\alpha=\beta = 1/2$ leads to the Jefferys prior whereas $\alpha = \beta = 1$ leads to the uniform prior). 
Finally, by a straightforward application of Bayes rule, the posterior distribution is found to be given by
\begin{equation}
Pr(p=x|s,f) = \frac{x^{s+\alpha-1}(1-x)^{f+\beta-1}}{B(\alpha+s, \beta+f)},\label{1.3}
\end{equation}
which is also a Beta distribution. In this way, as the experimenter provides the organism with additional pay-offs (one way or the other) the organism can infer a distribution about $p$, the probability of receiving $g$. Note that the mean of \eqref{1.3} is given by 
\begin{equation}
\tilde p = \frac{\frac{\alpha}{n}+\frac{s}{n}}{\frac{\alpha}{n}+\frac{\beta}{n}+1},\label{1.4}
\end{equation}
where $n = s+f$ is the total number of trials thus far. In this way, if the true value of $p$ is $\hat p$, then by giving the correction ratio of $s/n$ the organism can indeed infer the true value as $n$ gets large. Otherwise put
\begin{align}
\tilde{p}_\infty &= \lim_{n \to \infty} \frac{\frac{\alpha}{n}+\frac{s}{n}}{\frac{\alpha}{n}+\frac{\beta}{n}+1} ,\\
&= \hat p.
\end{align}
Observe that the prior information contained in $\alpha$ and $\beta$ is washed out (those terms go to zero), so if the experiment is carried out correctly and for long enough ($n \to \infty$), the prior beliefs of the organism do not matter.
\section*{Finite experiments}
In the previous section we showed that if the experiment is carried out \emph{ad infinitum} then the organism can make the correct inference.  With fixed rewards, they will therefore agree that the expected rewards of each arm are equal. However, experiments are necessarily finite. We now consider the implications of this.

Recall that in the experiments we are considering on one arm, from the point of view of the experimenter, a pay-off of $c$ is guaranteed. In other words $\hat{p}_1=1$. On the other arm, with probability $\hat{p}_2$ a reward of $a$ is given and probability $(1-\hat{p}_2)$ a reward of $b$. The experiment is designed such that 
\begin{equation}
c= \hat{p}_2a+(1-\hat{p}_2)b, \label{1.7}
\end{equation}
so that, with perfect knowledge (the frame of reference of the experimenter), there should be no preference for either arm if decisions are based solely on averages. 

However, from the point of view of the organism, at trial $n$ the estimated $\tilde{p}_{1}$ (setting $s=n$ in \eqref{1.4}) will be given by 
\begin{equation}
\tilde{p}_{1} = \frac{\frac{\alpha}{n}+1}{\frac{\alpha}{n}+\frac{\beta}{n}+1}, \label{1.8}
\end{equation}
 whereas $\tilde{p}_{2}$ will be estimated by 
\begin{equation}
\tilde{p}_{2}= \frac{\frac{\alpha}{n}+\frac{s}{n}}{\frac{\alpha}{n}+\frac{\beta}{n}+1},\label{1.9}
\end{equation}
assuming the same initial prior for both arms. Hence, for the organism to believe that the average pay-off on each arm is the same the following equality must hold:
\begin{equation}
c\tilde{p}_{1}= \tilde{p}_{2}a +(1-\tilde{p}_{2})b,\label{1.10}
\end{equation}
which, using \eqref{1.8}, \eqref{1.9} and simplifying is equivalent to
\begin{equation}
c= \left(\frac{\alpha +s}{\alpha+n}\right)a+ \left(\frac{\beta+n-s}{\alpha+n}\right)b.
\end{equation}
Using \eqref{1.7} on the left-hand side this is in turn equivalent to
\begin{equation}
\left(\hat{p}_2 - \left(\frac{\alpha +s}{\alpha+n}\right)\right)a+ \left(1-\hat{p}_2 -\left(\frac{\beta+n-s}{\alpha+n}\right)\right)b = 0.
\end{equation}
As $a$ and $b$ are, by design, greater than zero the above statement is only true if both cofactors are equal to zero. In particular, it must be that
\begin{equation}
\hat{p}_2 = \frac{\alpha +s}{\alpha+n}, \label{1.13}
\end{equation}
and 
\begin{equation}
\hat{p}_2 = \frac{\alpha - \beta+s}{\alpha+n}. \label{1.14}
\end{equation}
Note that both \eqref{1.13} and \eqref{1.14} can hold in only two ways. First, both may be true if $\beta = 0$. However, in this case the prior \eqref{1.2} is not a true distribution. When $\beta \to 0$ we can, in fact, interpret \eqref{1.2} as a Dirac delta function at $x=1$. Though in this case the entire problem is trivial as we no longer have any uncertainty. Second, $n$ may get arbitrarily large. However, in this section we are interested in precisely when this does not happen \emph{i.e.} when $n$ remains finite. In other words, in finite experiments \eqref{1.10} can never be true. The ramification of this is that even though the experimenter designs the experiment so that average pay-offs are equal, an organism that is performing these inferences will never agree in finite time. In this way, discussions of risk-preference to explain these broad group of experiments may be misleading.  

\section*{Discussion}
Hundreds of experiments have been designed such that given two choices the expected reward, but not the variance in reward, is equal on each choice \cite{kacelnik1996risky,shafir2000risk,weber2004predicting, hayden2009gambling,kawamori2010subjective}. This way experimenters attempt to infer risk preferences for organisms under a range of circumstances. If an organism consistently chooses the option that has less variance in reward they are deemed risk-averse. Conversely, if the organism chooses the option that has more variance they are deemed risk-prone. While risk-averseness occurs most frequently, risk-proneness has also been observed \cite{kacelnik1996risky}. Here, however, we have shown that these labels may be misleading. More importantly still, we have highlighted the importance of frames of reference in biology. In particular, we have stressed the potential pitfalls of studying behaviour without consideration of the point of view of the experimental subject. It makes, unfortunately, little sense to say that an organism will not make decisions based on expected rewards simply because we have programmed the experiment so that expected rewards are equal; there is no guarantee that the organism perceives said rewards to be equal. This is akin to the problems encountered in anthropomorphising animal behaviour whereby our own beliefs or motivations are projected onto non-human animals  \cite{wynne2004perils, dawkins2006through}.  

Bayesian approaches to understanding animal behaviour and in particular the use of statistical decision theory is of course nothing new \cite{mcnamara1980application, dall2005information,mcnamara2006bayes}. Most studies, rightly so, have focused on predicting decisions or phenotypes. More recently, others have considered the biological value of information itself \cite{mcnamara2009evolution,pike2016general, field2018ignorance}. Here, however, we focus instead on the limits of what a Bayesian organism can know before making decisions in the above experimental context. It is important to note that, unlike other studies, our work is divorced of any quantities to be optimised in order to make decisions such as utilities, reproductive values or energy budgets \cite{morgenstern1953theory,kahneman1979prospect, stephens1981logic,mishra2014decision,Sutton20118}. In this way, our work is about conflicting perceptions of the experimenter and the organism and therefore any inferences the experimenter can make about the organism, no matter what currency is optimised.  

In this paper, we have assumed that the organism in question is performing a Bayesian inference on the probability of receiving a certain reward. Starting with a general prior distribution, and sampling past rewards, we explicitly calculated a closed-form posterior distribution. A valid criticism, of course, is that organisms may not be behaving in a strictly Bayesian way. Indeed, much work has been done on this very question \cite{j2006animals, biernaskie2009bumblebees,louapre2011information}. However, as the Bayesian solutions are the optimal ones, we should expect natural selection to have moulded organisms that at least approximate Bayesian behaviour via so-called Rules of Thumb \cite{mcnamara1980application,mcnamara2006bayes,trimmer2011decision}. A further criticism, which pervades all of Bayesian analysis, is our choice of prior. Again we reiterate that it makes possible the calculation of a closed-form posterior. More importantly, however, we emphasise the versatility of the Beta distribution afforded by its hyperparameters $\alpha$ and $\beta$ which can control its concavity, skewness, symmetry and more.  Further,  for the particular values of $\alpha = \beta = 1/2$ and $\alpha = \beta= 1$ the Beta distribution reduces to the Jefferys and standard uniform distributions, respectively. We believe these two noninformative distibutions are particularly important for this study as, from the beginning of the experiment, there is no reason to suspect the organism has bias towards any initial value of $p$. For an extensive discussion of Bayesian prior choice, and in particular the use of noninformative priors, see Chapter 3 of Berger\cite{berger2013statistical}

With this set-up we found that while from the frame of reference of the experimenter the two choices have the same expected reward, any Bayesian organism will never agree in finite time. In light of this, we perhaps cannot appeal to differences in variance of reward to explain behaviour in these experiments. We are not implying then that organisms are only using averages to make decisions as in the elegant early work of Charnov \cite{charnov1976optimal,charnov1976optimalb}. Instead, we are pointing out that for these broad class of experiments we may not be able to appeal to variances so strongly as an explanatory variable. 

Though we have focussed on one class of experiments, our work points to the largely overlooked problem of frames of reference in studies of behaviour. If there is a mismatch between the beliefs of the experimenter and experimental subject, then we must be cautious to not draw conclusions based solely on our frame of reference. The first step, as taken here, is being conscious that such differences exist in the first place. For future work, it will be important to quantify just how divergent beliefs are and link this with existing work (such as in-built cognitive biases) on how organisms may in practice deal with these errors \cite{trimmer2011decision, johnson2013evolution}. For the particular experiments focused on here, it will be fruitful to consider explicit decision rules and currencies in order to generate \emph{in silico} data. Once done, it will be interesting to see if decisions based on expected rewards, variances or a combination of both most closely resembles the wealth of existing experimental data.  

\section*{Competing interests}
We have no competing interests.
\section*{Authors' contributions}
JMF carried out the research. JMF and MBB wrote the manuscript.
\section*{Funding}
JMF is funded by the Oxford University Charles Perkins Scholarship with additional financial support from UTS, Sydney.

\bibliographystyle{vancouver}

\begin{thebibliography}{10}

\bibitem{kacelnik1996risky}
Kacelnik A, Bateson M.
\newblock Risky theories—the effects of variance on foraging decisions.
\newblock American Zoologist. 1996;36(4):402--434.

\bibitem{shafir2000risk}
Shafir S.
\newblock Risk-sensitive foraging: the effect of relative variability.
\newblock Oikos. 2000;88(3):663--669.

\bibitem{weber2004predicting}
Weber EU, Shafir S, Blais AR.
\newblock Predicting risk sensitivity in humans and lower animals: risk as
  variance or coefficient of variation.
\newblock Psychological review. 2004;111(2):430.

\bibitem{parker1990optimality}
Parker GA, Smith JM.
\newblock Optimality theory in evolutionary biology.
\newblock Nature. 1990;348(6296):27.

\bibitem{markowitz1952portfolio}
Markowitz H.
\newblock Portfolio selection.
\newblock The journal of finance. 1952;7(1):77--91.

\bibitem{morgenstern1953theory}
Morgenstern O, Von~Neumann J.
\newblock Theory of games and economic behavior.
\newblock Princeton university press; 1953.

\bibitem{mcnamara1980application}
McNamara J, Houston A.
\newblock The application of statistical decision theory to animal behaviour.
\newblock Journal of Theoretical Biology. 1980;85(4):673--690.

\bibitem{stephens1981logic}
Stephens D.
\newblock The logic of risk-sensitive foraging preferences.
\newblock Animal Behaviour. 1981;29(2):628--629.

\bibitem{mcnamara2006bayes}
McNamara JM, Green RF, Olsson O.
\newblock Bayes’ theorem and its applications in animal behaviour.
\newblock Oikos. 2006;112(2):243--251.

\bibitem{pike2016general}
Pike RK, McNamara JM, Houston AI.
\newblock A general expression for the reproductive value of information.
\newblock Behavioral Ecology. 2016;p. 1296--1303.

\bibitem{field2018ignorance}
Field JM, Bonsall MB.
\newblock Ignorance can be evolutionarily beneficial.
\newblock Ecology and evolution. 2018;8(1):71--77.

\bibitem{Sutton20118}
Sutton NM, O’Dwyer JP.
\newblock Born to Run? Quantifying the Balance of Prior Bias and New
  Information in Prey Escape Decisions.
\newblock The American Naturalist. 2018;192(3):321--331.

\bibitem{hayden2009gambling}
Hayden BY, Platt ML.
\newblock Gambling for Gatorade: risk-sensitive decision making for fluid
  rewards in humans.
\newblock Animal cognition. 2009;12(1):201--207.

\bibitem{kawamori2010subjective}
Kawamori A, Matsushima T.
\newblock Subjective value of risky foods for individual domestic chicks: a
  hierarchical Bayesian model.
\newblock Animal cognition. 2010;13(3):431--441.

\bibitem{wynne2004perils}
Wynne CD.
\newblock The perils of anthropomorphism.
\newblock Nature. 2004;428(6983):606.

\bibitem{dawkins2006through}
Dawkins MS.
\newblock Through animal eyes: What behaviour tells us.
\newblock Applied Animal Behaviour Science. 2006;100(1-2):4--10.

\bibitem{dall2005information}
Dall SR, Giraldeau LA, Olsson O, McNamara JM, Stephens DW.
\newblock Information and its use by animals in evolutionary ecology.
\newblock Trends in ecology \& evolution. 2005;20(4):187--193.

\bibitem{mcnamara2009evolution}
McNamara JM, Stephens PA, Dall SR, Houston AI.
\newblock Evolution of trust and trustworthiness: social awareness favours
  personality differences.
\newblock Proceedings of the Royal Society of London B: Biological Sciences.
  2009;276(1657):605--613.

\bibitem{kahneman1979prospect}
Kahneman D.
\newblock Prospect theory: An analysis of decisions under risk.
\newblock Econometrica. 1979;47:278.

\bibitem{mishra2014decision}
Mishra S.
\newblock Decision-making under risk: Integrating perspectives from biology,
  economics, and psychology.
\newblock Personality and Social Psychology Review. 2014;18(3):280--307.

\bibitem{j2006animals}
J~Valone T.
\newblock Are animals capable of Bayesian updating? An empirical review.
\newblock Oikos. 2006;112(2):252--259.

\bibitem{biernaskie2009bumblebees}
Biernaskie JM, Walker SC, Gegear RJ.
\newblock Bumblebees learn to forage like Bayesians.
\newblock The American Naturalist. 2009;174(3):413--423.

\bibitem{louapre2011information}
Louapre P, Van~Baaren J, Pierre JS, Van~Alphen J.
\newblock Information gleaned and former patch quality determine foraging
  behavior of parasitic wasps.
\newblock Behavioral Ecology. 2011;p. 1064--1069.

\bibitem{trimmer2011decision}
Trimmer PC, Houston AI, Marshall JA, Mendl MT, Paul ES, McNamara JM.
\newblock Decision-making under uncertainty: biases and Bayesians.
\newblock Animal cognition. 2011;14(4):465--476.

\bibitem{berger2013statistical}
Berger JO.
\newblock Statistical decision theory and Bayesian analysis.
\newblock Springer Science \& Business Media; 2013.

\bibitem{charnov1976optimal}
Charnov EL.
\newblock Optimal foraging: attack strategy of a mantid.
\newblock The American Naturalist. 1976;110(971):141--151.

\bibitem{charnov1976optimalb}
Charnov EL.
\newblock Optimal foraging, the marginal value theorem.
\newblock Theoretical population biology. 1976;9(2):129.

\bibitem{johnson2013evolution}
Johnson DD, Blumstein DT, Fowler JH, Haselton MG.
\newblock The evolution of error: Error management, cognitive constraints, and
  adaptive decision-making biases.
\newblock Trends in ecology \& evolution. 2013;28(8):474--481.

\end{thebibliography}

\end{document}